\documentstyle[aps,twocolumn,epsf,floats]{revtex}

\def\kv{{\bf k}}

\begin{document}
\draft
\def\dbltopfraction{1.0}

\wideabs{
\title{Theoretical Model for Faraday Waves with Multiple-Frequency
Forcing}
\author{Ron Lifshitz and Dean M. Petrich}
\address{Condensed Matter Physics 114-36, California Institute of
Technology, Pasadena, CA 91125, USA}
\date{Recieved 4 April 1997}
\maketitle

\begin{abstract}
A simple generalization of the Swift-Hohenberg equation is proposed as
a model for the pattern-forming dynamics of a two-dimensional field
with two unstable length scales.  The equation is used to study the
dynamics of surface waves in a fluid driven by a linear combination of
two frequencies.  The model exhibits steady-state solutions with two-,
four-, six-, and twelve-fold symmetric patterns, similar to the
periodic and quasiperiodic patterns observed in recent experiments.
\end{abstract}
}

Parametric excitations of surface waves have been extensively studied
since their first discovery by Faraday~\cite{faraday} over a century
and a half ago.  In the basic experimental setup an open container of
fluid is subjected to vertical sinusoidal oscillations, which
periodically modulate the effective gravity. When the driving
amplitude $a$ exceeds a critical threshold $a_c$ a standing-wave
instability occurs with temporal frequency $\omega$ one half that of
the driving frequency. The characteristic spatial wavelength of the
standing-wave pattern is selected through the dispersion relation
$\omega(k)$ of the fluid. One typically observes patterns of stripes
or squares in such experiments. It is only in recent years that a
variety of additional patterns --- some with quasiperiodic rather than
periodic long range order --- have been
observed~\cite{danes,ef,gollub,fivefold,besson}. 
We shall focus here on a particular
set of experiments, performed by Edwards and Fauve~\cite{ef}, in which
a fluid was driven by a linear combination of {\it two\/}
frequencies, forming periodic patterns with 2-, 4-, and 6-fold
symmetry, and quasiperiodic patterns with 12-fold symmetry.

Previous theoretical
work~\cite{besson,malomed,zv,peilong,muller,newell} has focused mainly
on a description through amplitude equations with an angle-dependent
interaction $\beta(\theta_{ij})$ between pairs of modes. Such an
interaction, which is either postulated or derived from the underlying
microscopic dynamics, can be chosen to stabilize $N$-fold symmetric
patterns for arbitrary $N$. M\"uller~\cite{muller} has also used a set
of two coupled partial differential equations, where the pattern of a
primary field is stabilized by coupling to a secondary field which
provides an effective space-dependent forcing.  Newell and
Pomeau~\cite{newell} have coupled multiple fields in a similar way. In
both cases the coupling between the different fields is achieved
through resonant triad interactions, similar to the interactions we
shall introduce below.

We propose a simple rotationally-invariant model-equation, governing
the dynamics of a real field $u(x,y,t)$, which describes the amplitude
of the standing-wave pattern.  Our approach is different in that it
searches for the minimal requirements for reproducing the steady
states, which are observed in the experiments of Edwards and
Fauve~\cite{ef}. We incorporate into our model only the two most
essential aspects of the system:
\begin{enumerate}
\item The dynamics is damped at frequencies away from the two forcing
frequencies, and therefore the wavelengths involved in the selected pattern
lie in narrow bands about two critical wavelengths.
\item The driving used in the experiments is such that the up-down
symmetry, taking $u$ to $-u$, is broken allowing interactions among
triplets of standing plane-waves to exist. These triad interactions
are the {\it only\/} stabilizing mechanism for non-trivial patterns in our
rotationally-invariant model equation.
\end{enumerate} 
We capture the essential dynamics with a single field and without {\it
a priori\/} specifying any angle-dependent interactions among critical
modes. This allows for a meaningful comparison of the stability of
different $N$-fold symmetric states. We find patterns of 2-, 4-, 6-,
and 12-fold symmetry that are globally stable, but none with 8- or
10-fold symmetry, which is in agreement with the experimental
observations of Edwards and Fauve~\cite{ef}.

The supercritical instability of a homogeneous state to a striped
state is often modeled by the Swift-Hohenberg equation \cite{sheqn}
\begin{equation}
\partial_t u = \varepsilon u- (\nabla^2 + 1)^2 u - u^3,
\label{swift}
\end{equation}
which is variational,
\begin{equation}\label{relax}
\partial_t u = -\delta {\cal F} / \delta u,
\end{equation} 
driving the field $u(x,y,t)$ towards a minimum of the Lyapunov
functional (effective ``free energy'')  ---
\begin{equation}\label{free}
{\cal F} = \int \!dx\, dy\, \bigl\{- \frac12 \varepsilon u^2 + \frac12
[(\nabla^2+1)u]^2 + \frac14 u^4 \bigr\}.
\end{equation}
The first term in the Lyapunov functional (\ref{free}) favors the
growth of the instability whereas the quartic term is responsible for
its saturation by providing a lower bound for $\cal F$.  The growth
rate $\varepsilon$ of the instability is proportional to the reduced
driving amplitude $(a-a_c)/a_c$. The positive-definite gradient term
is small only near the critical wave number $k_c=1$, and thus
inhibits the growth of any instabilities with wave numbers away from
this value.

If the parametric forcing is such that the $u \rightarrow -u$ symmetry
is broken, then the Swift-Hohenberg ``free energy'' is modified by the
addition of a cubic term, $-\alpha u^3/3$.  Such a term allows triad
interactions of standing plane waves to lower the value of $\cal F$
and form hexagonal patterns.  The analysis of the Swift-Hohenberg
equation in the presence of this term is summarized, for example, in
the review by Cross and Hohenberg~\cite{ch}.  With single-frequency
forcing one cannot break the $u\to -u$ symmetry, but with certain
combinations of two frequencies the up-down symmetry is broken and
triad interactions become important.

We model the two-frequency parametric excitation of a fluid by
replacing the wavelength-selecting term in the Swift-Hohenberg
equation (\ref{swift}) by a similar term which damps out all modes
except those near one of {\it two\/} critical wavelengths:
\begin{equation}
\partial_t u = \varepsilon u - c(\nabla^2 + 1)^2(\nabla^2 +q^2)^2 u
+\alpha u^2 - u^3.
\label{lpeqn}
\end{equation}
The parameter $c$ can be scaled out, but we include it here because it
is used in the numerical simulations, shown later.  Other model
equations with similar wavelength-selection properties are possible.
We choose this equation because it is the simplest one that
incorporates the physics we are interested in --- it allows two
unstable length scales and contains triad interactions among the
different modes.  Since (\ref{lpeqn}) can be applied to any
pattern-forming system satisfying these requirements, it is not our
intention to provide a detailed derivation of it from any specific
underlying microscopic dynamics.

Let us turn now to an analytic investigation of the model equation
(\ref{lpeqn}).  When both $\varepsilon$ and $\alpha$ are sufficiently
small (or $c$ sufficiently large) the wavelength selection by the
gradient term is nearly perfect and the Lyapunov functional may be
written in Fourier space as
\begin{eqnarray}\label{fourier}
{\cal F} &= &-\frac12 \varepsilon \sum_{|\kv|=1,q} u_\kv u_{-\kv}
 -\frac13 \alpha \sum_{|\kv_i|=1,q} u_{\kv_1} u_{\kv_2}
 u_{-\kv_1-\kv_2} \nonumber \\
 & &\qquad + \frac14 \sum_{|\kv_i|=1,q} u_{\kv_1} u_{\kv_2} u_{\kv_3}
 u_{-\kv_1-\kv_2-\kv_3},
\end{eqnarray} where the summations  are restricted to wave vectors
whose magnitude is either 1 or $q$, lying on two rings in Fourier
space. The set of Fourier coefficients $u_\kv$, that give rise to the
lowest value of $\cal F$ for a given choice of the parameters
$\varepsilon$ and $\alpha$, determines the most favorable steady state
solution of the model equation (\ref{lpeqn}). We are only interested
in finding the global minimum of $\cal F$, thus establishing that our
model indeed predicts the existence of the patterns observed in the
two-frequency parametric forcing experiments.  Of course, this
approach may overlook meta-stable states or local minima of the free
energy. Note that with the omission of the gradient term, one may
perform a rescaling of the field $u\to\alpha u$.  The rescaled free
energy $\alpha^{-4}{\cal F}$ is then controlled by a single control
parameter 
\begin{equation}\label{epstar}
\varepsilon^\ast=\varepsilon/\alpha^2.
\end{equation}

\begin{figure}[t]
\begin{center}
\leavevmode
\hbox{%
\epsfxsize=4.0in
\epsffile{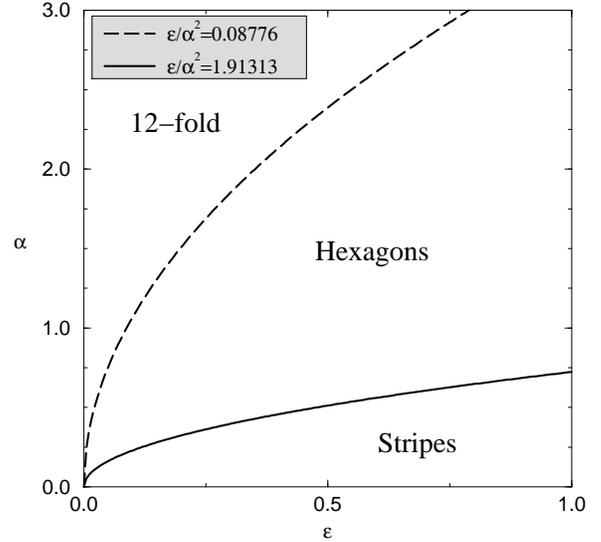}}
\end{center}
\caption{Phase diagram of the lowest-energy steady-state solutions of
the model equation (\ref{lpeqn}) for $q=2\cos(\pi/12)$. The phase
boundaries are lines of constant 
$\varepsilon^\ast=\varepsilon/\alpha^2$. }
\label{fig.phase}

\end{figure}

\begin{figure*}[t]
\begin{center}
\begin{picture}(440,390)
\put(-90,-220){\includegraphics{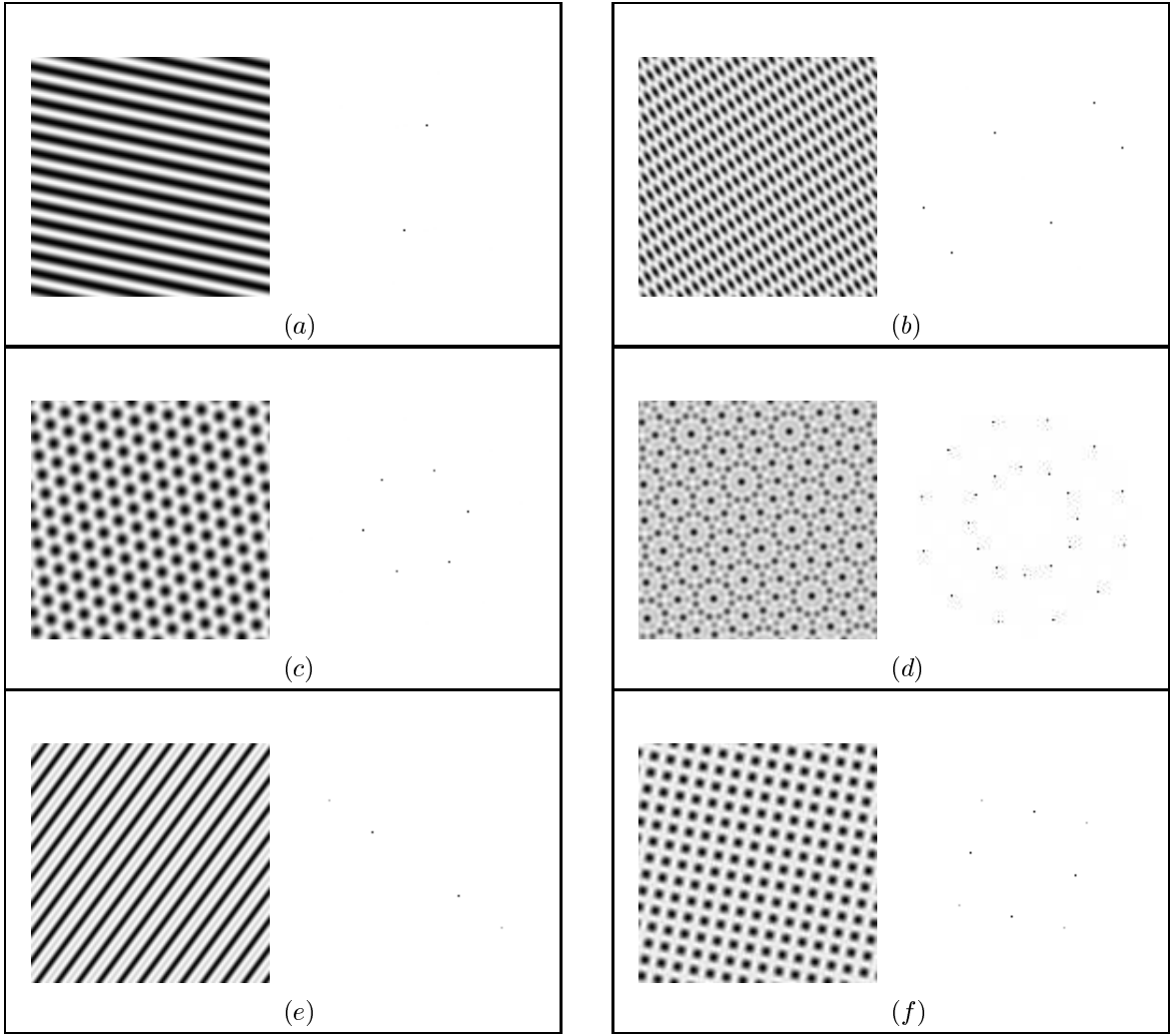}}

\end{picture}
\end{center}

\caption{Numerical solutions of the model equation (\ref{lpeqn})
showing real-space patterns along with their Fourier spectra for
different values of the control parameter $\varepsilon^\ast =
\varepsilon/\alpha^2$. The real-space images of $u(x,y,t\to\infty)$
show one quarter of the simulation cell with darker shades
corresponding to larger values of the field. All figures are drawn to
the same scale. In cases ($a$)-($d$) $q = 2\cos(\pi/12) =
(2+\protect\sqrt{3})^{1/2}$ : ($a$) A 2-fold pattern of stripes for
$\varepsilon^\ast =2$; ($b$) A 2-fold pattern of compressed hexagons
for $\varepsilon^\ast =0.1$; ($c$) A 6-fold pattern of perfect
hexagons for $\varepsilon^\ast =1.8$; ($d$) A 12-fold pattern for
$\varepsilon^\ast =0.015$.  In ($e$) $q=2$, $\varepsilon^\ast =0.04$,
yielding a 2-fold superstructure of stripes. In ($f$)
$q=\protect\sqrt{2}$, $\varepsilon^\ast =0.04$, giving rise to a
4-fold pattern of squares.  }
\label{fig.sim}

\end{figure*}

To study the formation of dodecagonal patterns we choose
$q=2\cos(\pi/12)$, which is the magnitude of the vector sum of two
unit vectors separated by an angle of 30 degrees. We minimize the
Lyapunov functional (\ref{fourier}) with respect to the Fourier
coefficients $u_\kv$ describing four different pattern candidates: (a)
a striped pattern with space group $P2mm$, whose Fourier spectrum
contains two opposite wave vectors of equal length; (b) a pattern of
perfect hexagons with space group $P6mm$, whose Fourier spectrum
contains a single 6-fold star of wave vectors; (c) a pattern of
compressed hexagons with space group $P2mm$, whose Fourier spectrum
contains four vectors on one ring and two vectors on the other ring;
and (d) a dodecagonal pattern with space group $P12mm$, whose Fourier
spectrum contains two 12-fold stars of wave vectors, one on each ring.

We use standard methods~\cite{mermin} to calculate $\cal F$ for each
of the cases. Because all the candidate patterns have symmorphic space
groups~\cite{sym} which are also centro-symmetric we may always take
all the Fourier coefficients on a given ring to be equal and their
phases may all be chosen such that they are either 0 or $\pi$. The
minimization of the Lyapunov functional is therefore always with
respect to no more than two real variables. We find the values of the
Lyapunov functional for the different patterns to be 
\begin{mathletters}\label{allfs}
\begin{eqnarray}
{\cal F}_2 &= &-\frac16 {\varepsilon^\ast}^2,
\label{f2} \\
{\cal F}_6 &= &{\cal F}_{4-2} = -\frac{4}{15^3}(1 + \sqrt{1+ 15
  {\varepsilon^\ast}}) \nonumber \\*
  &- &\frac{2}{15^2}(3 + 2\sqrt{1+ 15
  {\varepsilon^\ast}}){\varepsilon^\ast} - \frac{1}{10} {\varepsilon^\ast}^2,
\label{f6} \\
{\cal F}_{12} &= &- \biggl(\frac{10}{67}\biggr)^3(1 + \sqrt{1 +
  67{\varepsilon^\ast}/75}) \nonumber \\* 
  &- &\frac{20}{67^2} (1 +
  \frac23\sqrt{1 + 67{\varepsilon^\ast}/75}){\varepsilon^\ast} -
  \frac{9}{67}{\varepsilon^\ast}^2.
\label{f12}
\end{eqnarray} For  $\varepsilon^\ast>1.91313$
the striped pattern has the lowest free energy. For
$1.91313>\varepsilon^\ast>0.08776$ the 6-fold pattern of perfect
hexagons and the 2-fold pattern of compressed hexagons (denoted by
4-2), which are degenerate, are most favorable. For $\varepsilon^\ast<
0.08776$ the dodecagonal pattern is the most stable. These analytical
results are depicted in the phase diagram of Figure~\ref{fig.phase}.
Note that the phase diagram depicts only the boundaries between global
minima; in certain regions of the phase diagram additional states may
be locally stable.
\end{mathletters}

The model equation (\ref{lpeqn}), supplemented with periodic boundary
conditions, was solved numerically on a square domain using a
pseudo-spectral method.  The unit cell was typically chosen so that
the simulation region held about 30 wavelengths.  The simulation was
performed on a 256x256 grid, with Adams-Bashforth second-order
time-stepping. The value of $c$ was taken to be between 10 to 100.
Figures~\ref{fig.sim}($a$)-\ref{fig.sim}($d$) show the real-space and
Fourier-space results of the simulations with $q=2\cos(\pi/12)$ for
varying values of the control parameter $\varepsilon^\ast$.  The
results are consistent with the Lyapunov-functional analysis and the
phase diagram of Figure~\ref{fig.phase}.

Eight-fold and ten-fold symmetric patterns are not observed in our
model for any choice of $q$. An analytic calculation of the Lyapunov
functional (\ref{fourier}) for these patterns shows that it is greater
than the free energy ${\cal F}_6$ (\ref{f6}) of the six-fold state,
for any value of the control parameter $\varepsilon^\ast$.  This is in
accord with the experiments of Edwards and Fauve~\cite{ef}, where such
patterns are not observed. This does not rule out the possibility that
octagonal and decagonal patterns are locally stable but only that
within the limits of our model they are not globally stable.  Two
additional patterns that are observed in our model are a superposition
of stripes of periodicities $2\pi$ and $\pi$ (shown in
Figure~\ref{fig.sim}(e)) and a square pattern for $q={\sqrt{2}}$
(shown in Figure~\ref{fig.sim}(f)). The latter has been reported by
Edwards and Fauve. If one examines the Lyapunov functional
(\ref{fourier}) in its full generality by allowing the value of $q$
and all the amplitudes and phases to vary independently, other
patterns might be discovered. We have only examined the symmetric
patterns discussed here.

The simplicity of our model shows that for continuous media very
little is required to stabilize structures with quasiperiodic long
range order: two length scales and triad interactions. The reason that
12-fold patterns are stable and 8- and 10-fold patterns are not is
purely geometrical. In view of the Lyapunov functional
(\ref{fourier}), the crucial issue is the competition between the
number of modes, which tends to increase the value of $\cal F$, and
the number of triad interactions, which tends to decrease the value of
$\cal F$. The dodecagonal pattern of Figure~\ref{fig.sim}($d$)
contains 24 non-zero Fourier modes and {\it 32} distinct
triangles. The octagonal and decagonal patterns do not contain a
sufficient number of triangles to compete with the 6-fold pattern of
Figure~\ref{fig.sim}($c$). Our model confirms the conclusion of
Edwards and Fauve that ``12-fold patterns are more common than
previously supposed.''

Our simplistic model is clearly not adequate for studying the
structural stability quasicrystals in the solid state, yet it may
offer a very simple system in which to study general questions
regarding quasiperiodic order. These may include such questions as the
formation and propagation of defects and phase boundaries as well as
the dynamics of phason modes~\cite{muller2}. 
Moreover, we note that (\ref{lpeqn}) may apply to situations other
than Faraday waves.  Any physical system that can be tuned such
that two wavelengths undergo a simultaneous supercritical bifurcation
can be described by an equation similar to (\ref{lpeqn}).

An equation similar to (\ref{lpeqn}) could be used to study
multiple-frequency forcing of Faraday waves with more than just two
frequencies, as suggested by the title of this letter.  We may
speculate that with three or four forcing frequencies it might be
possible to stabilize quasiperiodic patterns with even higher orders
of symmetry, such as 18 or 24.  We leave the stability of higher-order
symmetric patterns as an open theoretical and experimental question.

We are grateful to Michael Cross, Jonathan Miller, and Peter Weichman
for many helpful discussions. This research was supported by the
California Institute of Technology through its Division Research
Fellowships in Theoretical Physics.

\end{document}